\documentstyle[preprint,aps]{revtex}

\begin{document}
\draft
\title{Charged exctions in the fractional quantum Hall regime}
\author{G. Yusa, H. Shtrikman, and I. Bar-Joseph}
\address{Department of Condensed Matter Physics, The Weizmann Institute of
Science, Rehovot 76100, Israel}
\maketitle

\begin{abstract}
We study the photoluminescence spectrum of a low density ($\nu <1$)
two-dimensional electron gas at high magnetic fields and low temperatures.
We find that the spectrum in the fractional quantum Hall regime can be
understood in terms of singlet and triplet charged-excitons. We show that
these spectral lines are sensitive probes for the electrons compressibility.
We identify the dark triplet charged-exciton and show that it is visible at
the spectrum at $T<2$ K. We find that its binding energy scales like $e^{2}/l
$, where $l$ is the magnetic length, and it crosses the singlet slightly
above 15 T.
\end{abstract}

\pacs{PACS: }

\narrowtext

The behavior of electrons in semiconductor heterostructures subjected to a
high magnetic field is governed by their mutual interactions. An important
tool, which has been intensively used for studying this behavior, is
photoluminescence (PL) spectroscopy. In a PL measurement an electron is
photo-excited from the valence to the conduction band. The hole, which is
created at the valence band, relaxes into the top of that band and
recombines with an electron from the Fermi sea. The interpretation of the
recombination spectrum is complicated, however, due to the presence of $e-h$
Coulomb interaction. The equality (up to a sign) of the $e-e$ and $e-h$
interactions may give rise to mutual cancellation of their contributions. It
was shown that in a symmetric system at the lowest Landau level this
cancellation renders all many-body effects invisible, and the only feature
that remains in the spectrum is the exciton \cite{MacDonald1}. This
symmetry, known as the hidden symmetry, is partially broken in realistic
systems. Experimental studies at the fractional quantum Hall (FQH) regime
revealed profound changes in the PL spectrum at fractional filling factors %
\cite{Turberfield,Goldberg,Kim}: the PL intensity exhibits strong minima or
maxima and new lines appear in the spectrum at the corresponding magnetic
fields. However, these changes could not be linked to a concrete physical
picture that takes into account the many-body $e-e$ and $e-h$ interactions
on equal footing.

An important development in the understanding of the behavior of the many
electron$+h$ system came through experimental studies of the PL of a dilute
two dimensional electron gas (2DEG) system. It was found that the ground
state of this system is the negatively charged exciton, $X^{-}$, which
consists of two electrons bound to a hole \cite{Kheng,Finkelstein1,Shields1}%
. It was shown that at zero magnetic field the two electrons are a
spin-singlet (${\bf J}=0$), and at high magnetic field another state, where
the electrons are a spin-triplet (${\bf J}=1$), becomes bound \cite%
{Shields2,Finkelstein2}. These observations triggered an intensive
experimental and theoretical work aimed at better understanding its
behavior, as it became clear that the $X^{-}$ may play an important role in
the PL of higher density 2DEG at high magnetic fields. Yet, there was a
significant qualitative discrepancy between the predicted behavior of these
bound states and the experimental results. It was argued that at the extreme
magnetic field limit the triplet should be the ground state of the system.
This is a manifestation of Hund's rule, which minimizes the repulsive
electrostatic energy of the electrons by having an anti-symmetric spatial
wavefunction. At zero magnetic field the Pauli exclusion principle sets an
energy price for the formation of such a state, hence the singlet is
preferred. However, at high magnetic fields a triplet state can be formed at
no cost of kinetic energy, since the two electrons can occupy degenerate
angular momentum states. Thus, a crossing behavior of the singlet and
triplet lines was predicted \cite{Palacios,Whittaker}. The experimental data
showed, however, a different behavior: the triplet binding energy was found
to rise and then saturate at a constant value \cite{Shields2,Finkelstein2}.
Surprisingly, no signature of singlet-triplet crossing was observed up to
very high fields \cite{Shields2,Hayne}. A solution to this disagreement was
recently proposed by Wojs {\it et al}. \cite{Wojs}. By calculating the
energy spectra of $2e+h$ states of a dilute 2DEG system it was found that 
{\it two different triplet states} are bound at high magnetic fields. These
states are distinguished by their total angular momentum $L$, one having $%
L=0 $ and the other $L=-1$. Consequently, the first could decay radiatively,
and was termed the ''bright'' triplet. The other state could decay only by a
scattering assisted process and was termed the ''dark'' triplet. It was
argued that the behavior observed experimentally is consistent with that of
the ''bright'' triplet. On the other hand, the ''dark'' triplet is the one
that crosses the singlet and becomes the ground state at high fields. Since
it can not recombine radiatively it should remain invisible. A recent PL
measurement at very high magnetic field (up to 40 T) showed evidences to the
existence of this dark state \cite{Munteanu}.

In this work we study the PL spectrum of a low density ($\nu <1$) 2DEG at
high magnetic fields ($B<15$ T) and low temperatures ($T=20$ mK). Our work
is motivated by recent theoretical studies that have suggested that the
charged exctions could be usefull in describing the PL spectrum of a 2DEG at
the FQH regime \cite{Palacios,Wojs}. Using a gated structure we are able to
follow the dependence of the PL spectrum on the filling factor, $\nu =en_{%
\text{e}}/hB$, not only by ramping the magnetic field at a constant density,
as is commonly done in PL experiments, but also by varying the density at
constant magnetic field. Our main finding is that the singlet and triplet
charged-exciton lines evolve {\it continuously} from the dilute limit into
the FQH regime, where they are sensitive probes for the many body
interactions. We identify conclusively the ''dark'' triplet and show that it
is visible at the spectrum at $T<2$ K. We find that its binding energy
scales like $e^{2}/l$, where $l$ is the magnetic length, and it should cross
the singlet slightly above 15 T.

The sample that we investigated is a single 20 nm GaAs/Al$_{0.37}$Ga$_{0.67}$%
As modulation doped quantum-well with electron mobility of $\sim 1\times
10^{6}$ cm$^{2}/$Vs. The MBE grown wafer is processed to a mesa structure
with a transparent gate electrode. The gated structure enables us to tune
the electron density $n_{\text{e}}$ continuously from $5\times 10^{9}$ to $%
2\times 10^{11}$ cm$^{-2}$. Most of our measurements were done in a dilution
refrigerator at a base temperature of $20$ mK, and a magnetic field of up to 
$15$ T that is applied along the growth direction of the wafer. The higher
temperature measurements ($T>1.5$ K) were done in a pumped He$^{4}$
cryostat. The sample was illuminated by Ti-sapphire laser with a photon
energy of $1.6$ eV and a power density of $2$ mW/cm$^{2}$. The PL was
collected using a fiber system and circular polarizers. All spectra shown in
this paper are at the $\sigma ^{-}$ circular polarization, in which a
spin-up electron from the lower Zeeman level recombines with a valence band
hole. The electron density under illumination is measured by finding the
values of the magnetic field $B$ that correspond to $\nu =1$ and $2$, where
drastic changes of the PL spectrum are observed \cite{Osborn}. The accuracy
of this method is better than $\sim 2\times 10^{9}$cm$^{-2}$. \ 

Figure 1 shows the PL spectrum at very low densities ($n_{\text{e}}\sim
5\times 10^{9}$ cm$^{-2}$) for the $\sigma ^{-}$ circular polarization. Let
us first focus on the temperature dependence of the spectrum, which is shown
in Fig. 1a for $9$ T. The spectrum at $4$ K is well studied and understood %
\cite{Shields2,Finkelstein2}: it consists of three main peaks associated
with the neutral exciton ($X^{0}$) and two charged-exciton peaks, labled as $%
X_{s}^{-}$ and $X_{t1}^{-}$. The two charged-exciton peaks are due to $e-h$
recombination from singlet or triple initial states, respectively. It is
clearly seen that as the temperature is decreased an additional peak,
labeled as $X_{t2}^{-}$, gradually appears between $X_{s}^{-}$ and $%
X_{t1}^{-}$, and becomes well resolved at $20$ mK. \ In the following we
show that $X_{t1}^{-}$ and $X_{t2}^{-}$ are the ''bright'' and ''dark''
triplets, respectively.

Figure 1b describes the evolution of the spectrum as the magnetic field is
varied between $0$ and $15$ T at $20$ mK. It is seen that at low fields ($%
B<4 $ T) the spectrum consists of only two peaks, the well known $%
X^{0}-X_{s}^{-} $ doublet \cite{Kheng,Finkelstein1}. This simple spectrum
changes at higher fields as two additional peaks, $X_{t1}^{-}$ and $%
X_{t2}^{-}$, split from the exciton and gradually shift to lower energies
with increasing magnetic field. Figure 2a summarizes the magnetic field
dependence of the peak energies. It can be clearly seen that both $X_{t1}^{-}
$ and $X_{t2}^{-}$ are unbound at zero magnetic field, and become bound at
some finite magnetic field. Examining their polarization properties we find
that both appear only at the $\sigma ^{-}$ polarization and do not have a
Zeeman-split counterpart. In Fig. 2b we show the binding energy of each
charged-exciton state, defined as its energy distance from $X^{0}$. It is
seen that the binding energies of $X_{s}^{-}$ and $X_{t1}^{-}$ exhibit a
rapid growth at low magnetic fields ($B<6$ T) and than saturate at a
constant value. This behavior was reported in several previous works \cite%
{Shields2,Finkelstein2}. The binding energy of $X_{t2}^{-}$, on the other
hand, grows monotonically with increasing the magnetic field and nearly
crosses that of the $X_{s}^{-}$ at $15$ T, consistent with the behavior
predicted for the ''dark'' triplet \cite{Whittaker,Wojs}. A quantitative
verification comes from the dependence of its binding energy on the magnetic
field. It can be seen that it is very well described by $0.1e^{2}/%
\varepsilon l$ (where $\varepsilon $ is the dielectric constant). This
functional dependence is indeed predicted for an ideal two-dimensional gas
in the lowest Landau level \cite{Palacios}, with a numerical coefficient of $%
0.0544$. The discrepancy in the coefficient is settled in theoretical
calculations that takes into account the finite well width and mixing with
higher Landau levels \cite{Whittaker,Wojs}. The magnetic field at which the
singlet-triplet crossing occurs, $\sim 15$ T, is, however, substantially
lower than predicted in these works ($30-40$ T). Very recent calculations
indicate that a slight displacement (of 0.5 nm) between the electron and
hole would shift this crossing magnetic field to the range observed in our
experiment \cite{Szlufarska}. Such a displacement might naturally occur in
our gated quantum well. We believe that this conclusive observation puts to
rest the debate over the triplet charged exciton. It should be noted that an
observation of the ''dark'' triplet was recently reported by Munteanu {\it %
et al.}, who have reinterpreted their previous high magnetic field
experiment on a high density 2DEG ($n_{\text{e}}=1.6\times 10^{11}$cm$^{-2}$%
) \cite{Munteanu}. However, the reported behavior at low fields is
inconsistent with that expected for a triplet charged exciton: Ref. \cite%
{Munteanu} shows a very large zero-field binding energy, while the triplet
is expected to be unbound.

Let us turn now to examine the dependence of the spectrum on filling factor.
In Fig. 3a we show the measured spectra as the density is changed from $%
1\times 10^{10}$ to $1.2\times 10^{11}$ cm$^{-2}$, at a constant magnetic
field of $10$ T. This density range corresponds to varying $\nu $ between $%
0.04$ to $0.5$. In Fig. 3b we present the peak energies of the neutral and
charged exciton lines as a function of $\nu $. It is seen that at as $\nu $
is increased from $0.04$ to $0.13$ the $X^{-}$ spectrum remains unchanged,
but the neutral exciton disappears. At this low density range the 2DEG is
most likely strongly localized, and does not form quantum Hall states. With
a further increase of the density the electron interactions become important
and the $X^{-}$ spectrum undergoes a drastic change: the two triplet lines, $%
X_{t1}^{-}$ and $X_{t2}^{-}$, gradually merge and at $\nu =1/3$ they form a
single strong peak. At $\nu >1/3$ this merged peak gradually weakens, until
it disappears from the spectrum at $\nu \approx 2/3$ (a weak recovery is
observed at $\nu =1$ ). The energy of the singlet state, on the other hand,
changes smoothly as we cross $\nu =1/3$, with no shift or casp. This
dependence on filling factor is general and is observed throughout the
magnetic field range, as demonstrated by the images of Fig. 4. Each
horizontal line in these images corresponds to a spectrum taken at a
different gate voltage, with the PL intensity being coded by colors. It is
seen that the energy separation between the lines vary with magnetic field,
but the merging of the two triplets at $\nu =1/3$ is clearly evident in all
the images. This observed behavior is in excellent quantitative and
qualitative agreement with that recently predicted by Wojs {\it et al.} \cite%
{Wojs}, who calculated{\it \ }the recombination energy and oscillator
strength of $e-X^{-}$ states of a low density 2DEG. These calculations
correctly predicts the energy dependence of the various lines, and in
particular - the merging of the two triplets at $\nu =1/3$ and the relative
insensitivity of the singlet state to $\nu $. The fact that one can
accurately obtain the PL spectrum around $\nu =1/3$ by considering the $%
e-X^{-}$\ interaction only is an important reassuring evidence for the
usefulness of the charged excitons in understanding the PL at the FQH
regime. 

We now wish to examine the dependence of the PL intensity on $\nu $ (Fig.
5). The behavior at $\nu <1/3$ is marked by a gradual decrease of $X_{t1}^{-}
$ intensity and a corresponding increase of $X_{t2}^{-}$ with increasing $%
\nu $. The higher density range is characterized by a more drastic
dependence of the PL intensity on $\nu $: the merged triplet exhibits strong
enhancement at exactly $\nu =1/3$ (and a weaker one at $2/5$),\ and the
singlet increases in a step-like fashion slightly above $\nu =1/3$. This
behavior is general and is seen throughout the magnetic field range: the
inset of Fig. 5 shows that the enhancements of the triplet occur
consistently at $\nu =1/3$ and $2/5$. Comparing this behavior with the
calculations of Ref. \cite{Wojs} we find a qualitative agreement at $\nu <1/3
$, but significant discrepancies around $1/3$. 

The interaction of the charged-exciton with the surrounding electrons is
sensitive to the compressibilty of the 2DEG. In the absence of the hole the
2DEG is spin polarized at the lowest Zeeman level and forms incompressible
states at fractional filling factors. The introduction of a positively
charged hole into the 2DEG creates a strong Coulomb attractive potential
near it, and the system minimizes its energy by creating a doubly occupied
bound state. In a dilute 2DEG ($\nu \ll 1/3$ ) this implies bringing two
electrons to the vicinity of the hole, forming either a spin-singlet or
triplet states. Earlier studies of the $D^{-}$ recombination in the presence
of a 2DEG have shown that this bound state is{\it \ }only weakly coupled{\it %
\ }to the rest of the electrons: the quasi-hole that is formed at the lowest
Zeeman level tends to migrate to the vicinity of the electron pair, while
the remaining electrons move away into larger orbits \cite{Jiang}. Thus, the
bound state is an $X^{-}$ that is effectively isolated from the 2DEG, as
manifested in the spectra near $\nu =0$. As we approach $\nu =1/3$
interactions with the rest of the electrons and the formation of an
incompressible state have to be taken into account. The enhancement of the
triplet PL intensity at $1/3$ and $2/5$ is, therefore, surprising: as
explained above, the triplet has a non-vanishing total angular momentum,
hence it can recombine through scattering assisted processes only. The
incompressibilty of the 2DEG at $\nu =1/3$ should suppress $e-e$ scattering,
and thus, give rise to a reduction in the triplet intensity. A possible
explanation is that the incompressiblity of the 2DEG suppresses also its
effectivity in screening the remote donors potential, and the fluctuations
at the 2DEG plane grow. This fluctuating potential can scatter the triplet
charged excitons and enable their recombination.  It should be noted that
the triplet state is not dark throughout the filling factor range: its
intensity is significantly higher than calculated and is nearly the same as
that of the singlet. This implies that scattering processes can efficiently
transfer its excess angular momentum and enable it to recombine radiatively.
The fluctuating donors potential could explain this behavior as well.

We wish to acknowledge very fruitfull discussion with A. Stern and A Esser.
Special thanks to D. Sprinzak and Y. Ji for their assistance in the dilution
refrigerator. This research was supported by the Minerva Foundation.

\figure Fig. 1: (a) The PL spectrum at at low electron density, $n_{\text{e}%
}\sim 5\times 10^{9}$ cm$^{-2}$, as a function of temperature. (b) The PL\
spectrum at 20 mK as a function of magnetic field.

\figure Fig. 2: (a) The peak energies and (b) binding energies of a dilute
2DEG ($n\sim 5\times 10^{9}$ cm$^{-2}$) as a function of $B$.

\figure Fig. 3: (a) The PL spectra at 10 T for $0.04<\nu <0.50$. (b) The
peak energies as a function of $\nu $.

\figure Fig. 4: Contour plots of the PL spectra as a function of $\nu $ at
different magnetic fields, (a) $B=8$ T, (b) $11$ T, and (c) $13.5$ T. The PL
intensity is color coded, such that blue is low and red is high.

\figure Fig. 5: The PL peak intensity of the three charged-exciton lines as
a function of $\nu $. Inset: The value of $\nu $ where the PL intensity of $%
X_{t2}^{-}$ is enhanced as a function of magetic field.

\end{document}